\documentclass{IEEEtran}
\usepackage{cite}

\usepackage{amsmath,amssymb,amsfonts}
\usepackage{graphicx}
\usepackage{textcomp,nicefrac}
\usepackage[switch]{lineno}
\def\BibTeX{{\rm B\kern-.05em{\sc i\kern-.025em b}\kern-.08em
T\kern-.1667em\lower.7ex\hbox{E}\kern-.125emX}}
\begin{document}
\title{The  study of 4H-SiC LGAD after proton radiation}
\author{Sen Zhao, Jiaqi Zhou, Chenxi Fu, Congcong Wang, Suyu Xiao, Xinbo Zou, Haolan Qv,Jiaxiang Chen, Xiyuan Zhang and Xin Shi 
\thanks{This work is supported by the National Natural Science Foundation of China (Nos.~12205321, 12375184, 12305207, and 12405219), National Key Research and Development Program of China under Grant No. 2023YFA1605902 from the Ministry of Science and Technology, China Postdoctoral Science Foundation (2022M710085), Natural Science Foundation of Shandong Province Youth Fund (ZR2022QA098) under CERN RD50-2023-11 Collaboration framework. (Corresponding authors:~Xin Shi; Xiyuan Zhang.)}
\thanks{Sen Zhao is now with the Institute of High Energy Physics, Chinese Academy of Sciences, Beijing
100049, China.}
\thanks{Jiaqi Zhou is now with Jilin University, Changchun 130015, China, and also with the Institute of High Energy Physics, Chinese Academy of Sciences, Beijing 100049, China.}
\thanks{Congcong Wang and Chenxi Fu are with the Institute of High Energy Physics, Chinese Academy of Sciences, Beijing 100049, China.}
\thanks{Suyu Xiao is with Shandong Institute of Advanced Technology, Jinan 250100,
 China.}
 \thanks{Xinbo Zou, Haolan Qv and Jiaxiang Chen are with ShanghaiTech University, Shanghai 201210, China.}
\thanks{Xin Shi and Xiyuan Zhang are with
the Institute of High Energy Physics, Chinese Academy of Sciences, Beijing
100049, China. State Key Laboratory of Particle Detection and Electronics, Beijing 100049, China. (e-mail: zhangxiyuan@ihep.ac.cn,
shixin@ihep.ac.cn).}}
\maketitle

\begin{abstract}
Silicon carbide (SiC) is a promising material for radiation monitoring in harsh environments, due to its low dark current, high breakdown voltage, high thermal conductivity, and radiation hardness.~This work investigates a SiC-based Low-Gain Avalanche Detector (LGAD), named SICAR, with a gain factor of~2 to 3, under 80 MeV proton irradiation up to $1\times 10^{14}$~$n_{eq}/cm^{2}$. Electrical characterization via I-V, C-V, and $\alpha$ particle injection reveals an increase in threshold voltage and a 2 to 4 order of magnitude reduction in leakage current, while charge collection efficiency decreases by about 50\%. X-ray diffraction (XRD) and capacitance deep-level transient spectroscopy (C-DLTS) were employed to characterize the lattice structure and deep-level defects before and after irradiation. Deep-level defect characteristics were integrated into TCAD simulations to develop an electrical degradation model for SiC LGADs. A linear defect-flux relationship is established in the model, showing agreement with experimental results. 

\end{abstract}


\section{INTRODUCTION}
\label{sec:introduction}

\IEEEPARstart{S}{emiconductor} radiation detectors are widely used in high-energy physics, space exploration, medical imaging, nuclear monitoring, and security systems. In high-energy physics, to explore the energy frontier, both Chinese and European particle physicists  have proposed plans for future 100-km-circumference circular colliders\cite{CEPCStudyGroup:2018rmc}. These proton-proton colliders would reach 100~TeV, nearly an order of magnitude higher than the HL-LHC's energy. However, such high-energy collisions entail harsher radiation environments and higher pileup rates, posing unprecedented challenges to detector radiation hardness.~Moreover, while detector operation in intense radiation fields imposes stringent cooling requirements, room-temperature-stable detector technologies are emerging as a new frontier for future development\cite{Radiationharness,Radiationharness2,Radiationharness3}. \par
Silicon carbide (SiC), as a wide-band gap semiconductor, possesses a high breakdown electric field, excellent thermal conductivity, superior carrier mobility, and high saturation drift velocity\cite{SELLIN2006479,Harley-Trochimczyk_2017,WOS:000803113800029}. These properties enable its long-term stable operation under both room-temperature and intense-radiation environments. Although the wide band gap of the SiC detector greatly reduces thermal carrier generation and consequently suppresses noise, it exhibits a non-negligible disadvantage that for a given particle energy (assuming full energy conversion to electron-hole pairs), SiC produces only one-third of the electron-hole pairs, resulting in a lower pulse amplitude compared to silicon. Thus, the development of SiC Low Gain Avalanche Diodes (LGAD) and research on their radiation-resistant properties merit thorough investigation.\par

Previous studies indicate that SiC detectors (e.g., PIN and Schottky diodes) exhibit excellent radiation resistance to gamma/electron irradiation, but suffer severe charge collection efficiency (CCE) degradation (>80\%) under 24 GeV proton irradiation at an equivalent neutron dose of $2.5\times 10^{15}$\cite{PINRA}. In contrast to conventional PIN/Schottky devices, the inherent low-gain characteristic of SiC LGADs enables partial recovery of charge collection efficiency (CCE) post-irradiation. This unique property necessitates investigation of defect-mediated degradation mechanisms and physical models in SiC LGADs.

This study characterizes the our first-generation 4H-SiC LGAD (SICAR1) under 80 MeV proton irradiation~($2\times 10^{11}$ - $1\times 10^{14}$ n$_{eq}/cm^{2}$~equivalent neutron flux), with systematic evaluation of I-V, C-V, and CCE properties.~To investigate the degradation mechanisms induced by proton irradiation, we characterized the lattice properties~(via XRD) and deep-level defects~(via C-DLTS) in fully epitaxial (ion-implantation-free) 4H-SiC device structures before and after irradiation.~Electrical and defect analysis revealed dominant proton radiation-induced defects in SiC LGADs, enabling development of a dose-dependent defect model validated by experimental results.~This work demonstrates the proton radiation tolerance of SiC LGADs and establishes a proton irradiation-defect correlation model through combined experimental and simulation approaches, guiding future device optimization and radiation studies.

\section{Methods and materials}

This study investigates the degradation mechanisms of 4H-SiC LGAD diodes under proton irradiation. The device structure comprises: (i) a 50-$\mu$m-thick N-type epitaxial layer (doping concentration: $3.8 \times 10^{14}$ cm$^{-3}$), (ii) a 1-$\mu$m-thick N-type gain layer ($8.6 \times 10^{16}$ cm$^{-3}$), (iii) a heavily doped P-type epitaxial anode layer, and (iv) an N-type conductive substrate cathode with <0001> $\pm$4$^{o}$ off-cut orientation. The device etching angle ranges from 60 to 90 degrees, with an etching depth of 1.6 micrometers. As the first-generation 4H-SiC LGAD developed by our group, its detailed structural parameters can be seen from~\figurename~\ref{fig1}~(a)\cite{P9}. \par

Proton irradiation experiments were performed on the devices using an 80 MeV proton beam at the China Spallation Neutron Source (CSNS, Dongguan). The beam was incident perpendicular to the device surface with flux of $2\times 10^{11}$~$n_{eq}/cm^{2}$, $3.5\times 10^{13}$~$ n_{eq}/cm^{2}$ and $1\times 10^{14}$~$ n_{eq}/cm^{2}$. Post-irradiation testing was conducted in the 20~$^{o}$C environment.\par

Current-voltage (I-V) characteristics were measured using a Keithley 2470 source meter on a probe station. Capacitance-voltage (C-V) characterization was performed at 100 kHz using a Keysight E4980A Precision LCR meter with an external bias adapter. The lattice structures were investigated by X-ray diffraction (XRD, high-resolution Panalytical Empyrean, K$\alpha$ radiation) and the deep-level defects were characterized by deep-level transient spectroscopy (DLTS) measurement utilizing the FT 1230 HERA DLTS system. Device simulations were performed using Synopsys Sentaurus TCAD tools.  \par

The charge collection system comprises a $^{241}$Am $\alpha$-particle source and a single-channel readout board featuring a transimpedance amplifier (TIA), powered by a Keithley 2470 high-voltage supply and GWINSTEK GPD-3303S low-voltage unit, with signal acquisition performed using a Tektronix DPO-7354C oscilloscope (2.5 GHz bandwidth).

\section{Results and Discussion}
\par
\begin{figure*}[!t]
    \centerline{\includegraphics[width=7in]{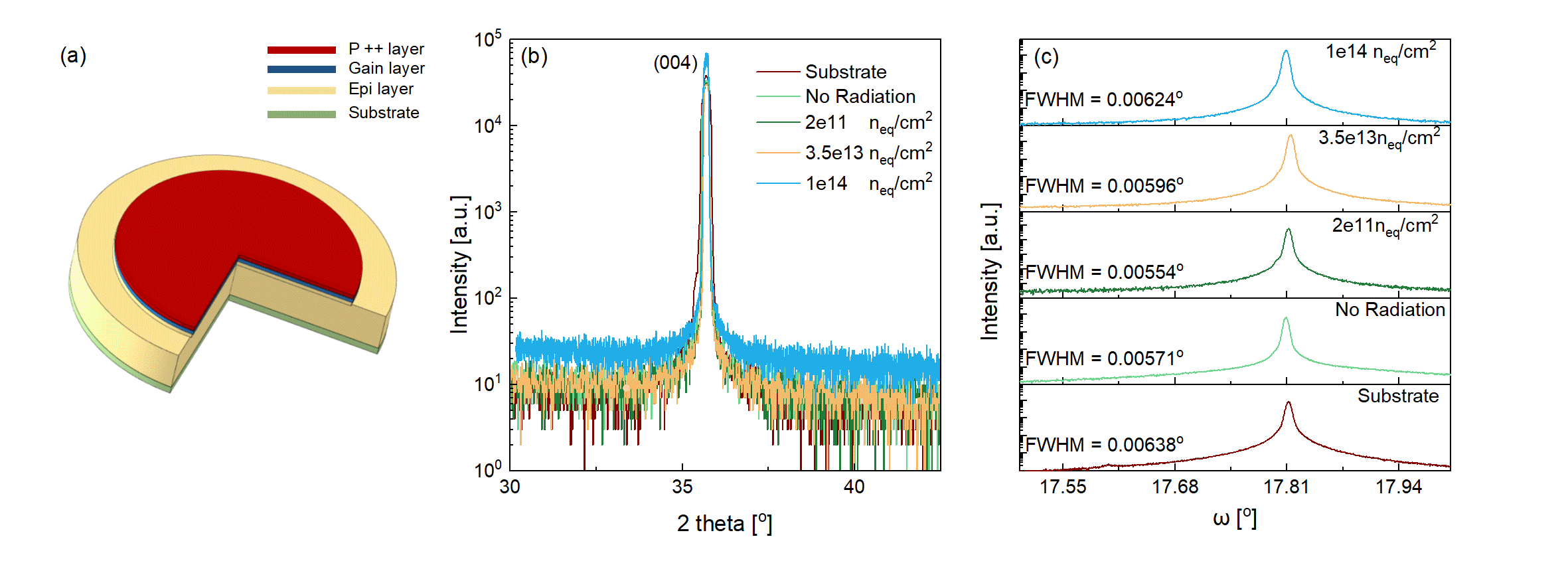}}
     \caption{(a)~Device epitaxial structure cross-section.~(b)~2$\theta$ scan result of all samples.~(c)~Rocking Curve result of all samples.}
    
    \label{fig1}
\end{figure*}

As shown in \figurename~\ref{fig1} (a), the 4H-SiC LGAD was fabricated using a full epitaxial growth approach, which enables precise control of the thickness and graded doping profiles in the n-type active region, gain layer, and P++ layer. The method used in 4H-SiC LGAD eliminates lattice damage typically induced by conventional high-temperature ion implantation and annealing processes\cite{ion-implant}.~In 4H-SiC, increasing ion implantation induces progressive lattice mismatch ($\Delta$C/C$_0$ > 0.5\% at $1\times 10^{20}~cm^{-3}$) and defect formation\cite{10.1063/1.4720435}. The epitaxial N-doped gain and Al-doped P++ layers maintain well lattice coherence with the SiC substrate, in contrast to the lattice mismatch and crystallographic tilt induced by high-dose ion-implantation into 4H-SiC. \par

This is confirmed by the XRD $\theta$-2$\theta$ spectrum (\figurename~\ref{fig1} (b)),  which reveals identical (004) peak positions for both epitaxial layers and substrate. In addition, the epitaxial-grown device maintains crystallographic stability even under proton irradiation ($2\times 10^{11}$–$1\times 10^{14}$~$ n_{eq}/cm^{2}$).To further assess the crystalline quality and structural integrity of the epitaxial layer at varying irradiation doses, \figurename~\ref{fig1} (c) shows the corresponding rocking curves. It can be observed that the full width at half maximum (FWHM) of the epitaxial layer is consistent with that of the substrate, exhibiting good peak symmetry and exceptionally narrow FWHM measurements below 0.006$^{o}$, which confirms a well-defined crystal structure.~In summary, proton irradiation (80 MeV, $2\times 10^{11}$–$1\times 10^{14}$~$ n_{eq}/cm^{2}$) demonstrates negligible impact on the lattice quality of the fully epitaxial-grown 4H-SiC LGAD.\par

\begin{figure*}[h]
    \centerline{\includegraphics[width=6in]{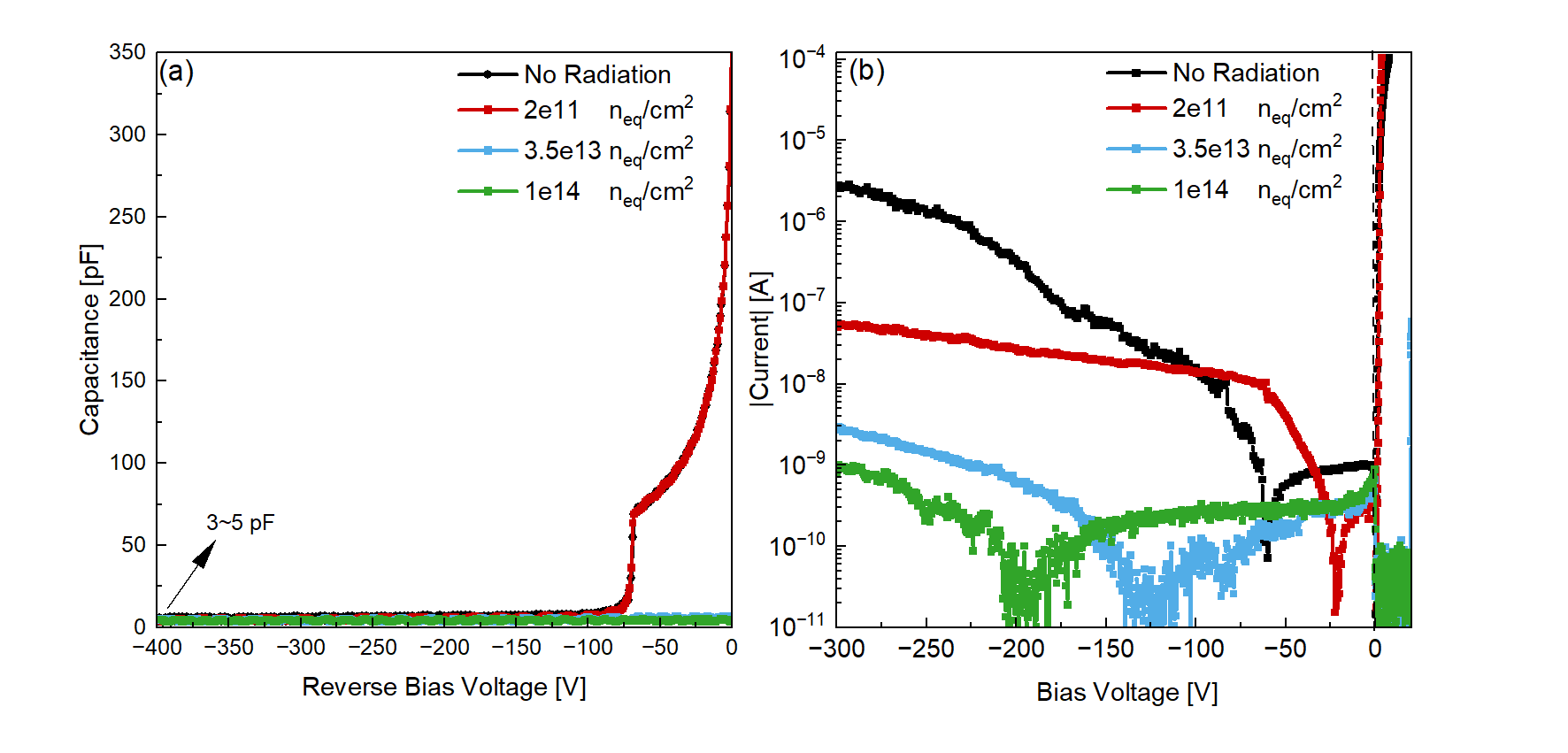}}

   \caption{(a) The capacitance vs reverse bias voltage of radiated samples. (b) The current vs bias voltage of radiated samples}
    \label{fig2}
\end{figure*}

 The capacitance-voltage (C-V) characteristics of LGAD detectors before and after irradiation are shown as \figurename~\ref{fig2} (a). The test system exhibited a parasitic capacitance of 3-5 pF. The C-V curves demonstrate that both unirradiated and low-flux ($2\times 10^{11}~n_{eq}/cm^{2}$) samples show a distinct step transition at 70~V bias, indicating complete depletion of the gain layer.
 At high irradiation flux ($ \geq 3.5\times 10^{13}~n_{eq}/cm^{2}$), the devices exhibit significant degradation in their capacitance-voltage characteristics, manifested by voltage-independent depletion depth\cite{flatcv}.~This phenomenon stems from strong compensation effects in the drift region, resulting in substantial reduction or even complete elimination of the effective doping concentration\cite{compensation1}. \par
 
The current-voltage (I-V) characteristics of LGAD devices before and after irradiation are presented as \figurename~\ref{fig2} (b). The observed evolution of both reverse and forward leakage currents with increasing irradiation flux demonstrates systematic degradation in the device's rectification behavior. Notably, the reverse leakage current demonstrates a remarkable reduction of 2 to 4 orders of magnitude with increasing irradiation flux. This behavior stands in sharp contrast to conventional silicon~(Si) detectors, which typically exhibit a rapid increase in leakage current under identical irradiation conditions \cite{Si_radiation}. Furthermore, the forward leakage current exhibits a flux-dependent decrease, with the turn-on threshold voltage increasing proportionally to irradiation flux. At the maximum tested flux ($1\times 10^{14}$ $n_{eq}/cm^{2}$), the device's forward turn-on threshold voltage exceeds 40 V.~These behaviors are associated with conduction resistance increase, due to the generation of radiation-induced defects\cite{ForwardIV}. Preliminary defect characterization results indicate that internal deep energy level defects may be a contributing factor to the observed phenomena\cite{IV-DLD,IV-DLD2}.\par

In order to investigate the degradation of electrical properties, the presence of deep energy level defects has also been detected. To investigate the intrinsic and proton irradiation-induced ($2\times 10^{11}$~$ n_{eq}/cm^{2}$) deep-level defects in 4H-SiC LGAD devices, we conducted comparative C-DLTS measurements on both non-irradiated and irradiated samples. Since C-DLTS relies on the transient response of junction capacitance to bias voltage variations, this technique becomes ineffective when the device capacitance exhibits voltage independence (i.e., $\frac{dC}{dV} \approx 0$). Our study reveals that when the proton irradiation flux exceeds $3.5\times 10^{13}$~$ n_{eq}/cm^{2}$, the C-V characteristics degrade into a constant capacitance plateau, preventing the detection of thermal emission signals from deep-level defects via conventional capacitance transient measurements. Consequently, C-DLTS analysis is no longer feasible under such conditions.

\begin{figure*}[h]
    \centerline{\includegraphics[width=6.5in]{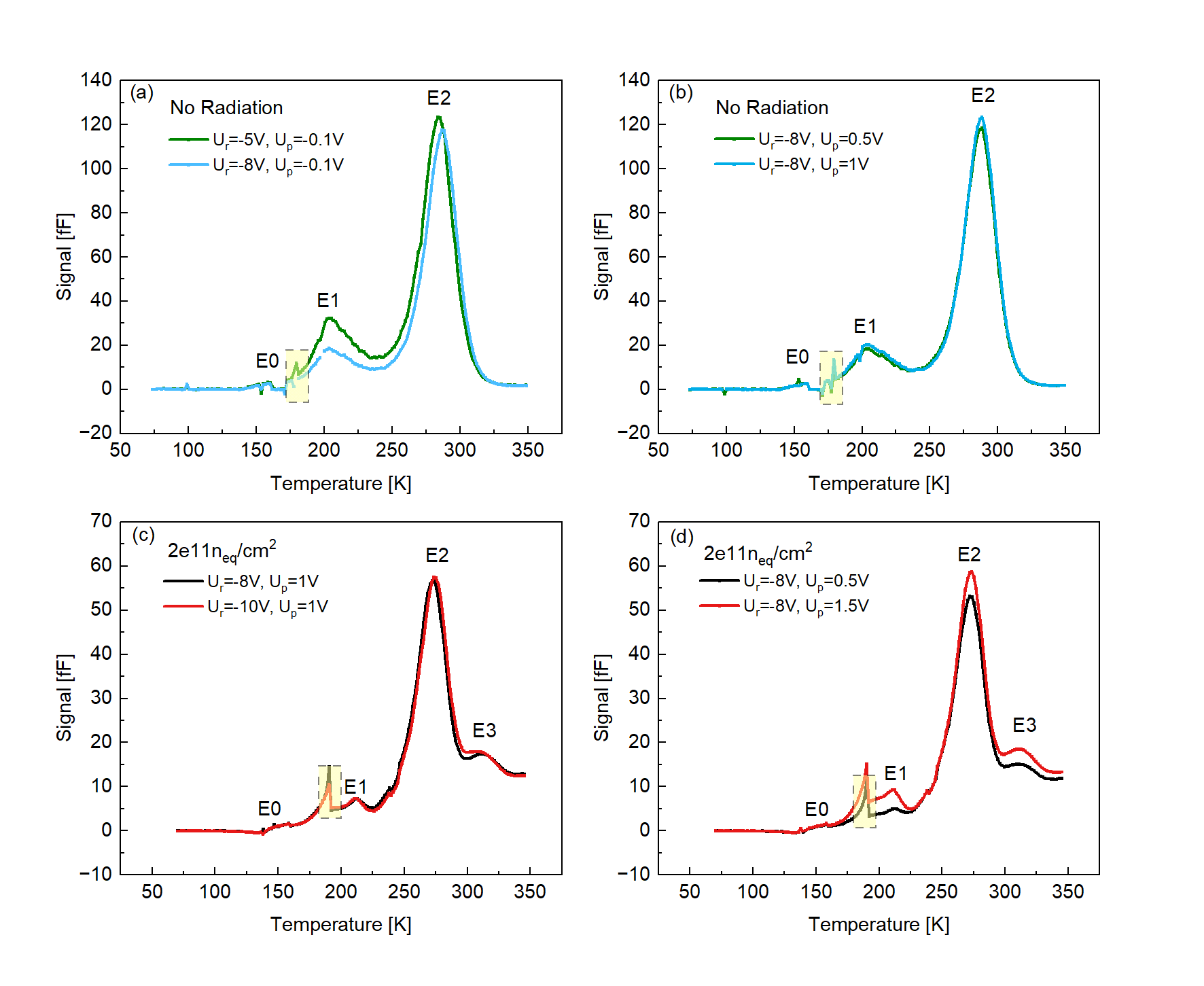}}
    \caption{DLTS spectra of no radiation and $2\times 10^{11}~n_{eq}/cm^{2}$ 4H-SiC sample. The bias voltage of the device is kept constant and the pulse voltage is varied to observe the ability of the defects to respond to the electric field.((a) No radiation sample. U$_p$ is unchanged, changing U$_r$. (c) $2\times 10^{11}~n_{eq}/cm^{2}$  sample. U$_p$ is unchanged, changing U$_r$.) The constant pulse voltage of the device changes the bias voltage to observe the concentration of defects. ((b) No radiation sample. U$_r$ is unchanged, changing U$_p$.  (d) $2\times10^{11} ~n_{eq}/cm^{2}$  sample. U$_r$ is unchanged, changing U$_p$.)}
    \label{fig3}
\end{figure*}

The DLTS test result of the no radiation samples are shown in \figurename~\ref{fig3} (a) and (b). The profile of defects is shown in Table~\eqref{tab4}.~Three defects were discovered, labeled by E0,E1 and E2, respectively. The E0 defect is defined as the formation of a shallow energy level defect from the metal, with a limited effect on device performance\cite{E0}.\par
 Under fixed pulse voltage conditions, bias variation induces distinct responses in the C-DLTS spectrum: the E2 defect peak shifts toward higher temperatures while the E1 peak remains stable. This contrast reveals that the E2 defect is highly sensitive to electric fields, and its energy level position or carrier trapping/emission kinetics may be significantly affected by the applied electric field. Specifically, the change in bias may have altered the local electric field distribution around the defect, which in turn modulated the energy level position or trapping cross-section of the defect, resulting in a shift of its characteristic peaks in the C-DLTS spectrum.In contrast, the E1 defect exhibits a stable energy level position upon bias change, suggesting that the defect presents an electrically neutral state after its carrier emission. This property implies that the energy level positions of the E1 defect are more fixed in the bandgap and are not significantly affected by the applied electric field. The electroneutral behavior of the E1 defect further suggests that it does not introduce additional charge state changes during the carrier emission process, and thus has less influence on the electric field distribution. ~It is reported that E1 and E2 are the EH1 ($E_C$-0.44eV) and $Z_{1,2}$ ($E_C$-0.67eV),respectively\cite{DLTS,Z12}. The sources of the two traps are probably $V_{Si}(3-/=)$ and $V_C(=/0)$\cite{EH1/2,originZ121,originZ122}.\par

\begin{table*}[t]
\caption{{Trap details in the no radiation sample}}
\label{table-dlts-nora}
\setlength{\tabcolsep}{4pt}
\centering
\begin{tabular}{cccc}
\hline
Trap & Activation energy [eV] &Capture cross-section [cm$^{2}$]&Trap concentration [cm$^{-3}$] \\
\hline
E0 &0.13&$1.88\times 10^{-14}$ &$1.01\times10^{12}$\\E1 &0.44&$1.88\times 10^{-14}$ &$1.01\times10^{12}$\\E2 &0.63&$1.07\times 10^{-14}$ &$6.74\times10^{12}$\\
\hline
\end{tabular}
\label{tab4}
\end{table*}
The DLTS test result of the $2\times 10^{11} n_{eq}/cm^{2}$ samples are shown in \figurename~\ref{fig3} (c) and (d). The profile of defects is shown in Table~\eqref{tab5}. The three defects E0, E1, and E2 exhibited characteristics analogous to the DLTS test results of the no radiation device, thereby being classified as intrinsic defects of the 4H-SiC device. A novel defect, designated E3, was identified. ~It can be seen that the irradiation-induced E3 defects show low sensitivity to changes in the applied electric field. This insensitivity indicates that their energy level positions and trapping cross sections remain stable under different bias conditions.
According to established studies \cite{DLTS}, the E3 defect is conclusively identified as the EH3 center with an energy level at ($E_C$-0.70 eV). This defect is associated with the doubly negative charge state ($V_{Si}$) of silicon vacancies \cite{EH3}.\par


\begin{table*}[t]
\caption{{Trap details in the flux of 2$\times$ 10$^{11}~n_{eq}/cm^{2}$ sample}}
\label{table-dlts-2e11}
\setlength{\tabcolsep}{4pt}
\centering
\begin{tabular}{cccc}
\hline
Trap & Activation energy [eV] &Capture cross-section [cm$^{2}$]&Trap concentration [cm$^{-3}$] \\
\hline
E0 &0.13&$1.88\times 10^{-14}$ &$1.01\times10^{12}$\\E1 &0.46&$1.66\times 10^{-15}$ &$1.38\times10^{12}$\\E2 &0.61&$1.27\times 10^{-15}$ &$1.07\times10^{13}$\\E3 &0.70&$1.30\times 10^{-15}$ &$3.48\times10^{12}$\\
\hline
\end{tabular}
\label{tab5}
\end{table*}



Due to the constraints imposed by the experimental conditions, defects manifesting at elevated temperatures remained undetected. However, the defect EH6,7($E_c$-1.6eV), which manifests at elevated temperatures, is a significant concern within the 4H-SiC material\cite{EH65_LEVEL}. The defect EH6,7 was incorporated into the model through the utilization of simulations. Given its provenance from carbon vacancies, akin to Z$_{1,2}$, it is assumed that the defects are generated at a rate consistent with that observed in Z$_{1,2}$\cite{ORIGIN_EH67}. In general, the concentration of EH6,7 inside the device is approximately $1 \times 10^{12}~cm^{-3}$\cite{EH67_CON}. In simulation, the concentration is set to be $6.74 \times 10^{12}~cm^{-3}$  and the electron capture cross section is about $2 \times 10^{-13}~cm^{-2}$ \cite{EH67_CON}. \par
The results of defect characterization hypothesized that the defect concentration and irradiation flux demonstrate a positive relationship, and that the scale factor is the defect generation rate. The following defect concentration distribution can be estimated by  $$n_{defects}  = n_0 + \eta \times flux$$ as in silicon, where $n_0$ is the intrinsic defect concentration, $\eta$ is the defect generation rate, and flux is the radiation dose.\cite{NIELcom}\par

\begin{figure*}[!t]
    \centerline{\includegraphics[width=8in]{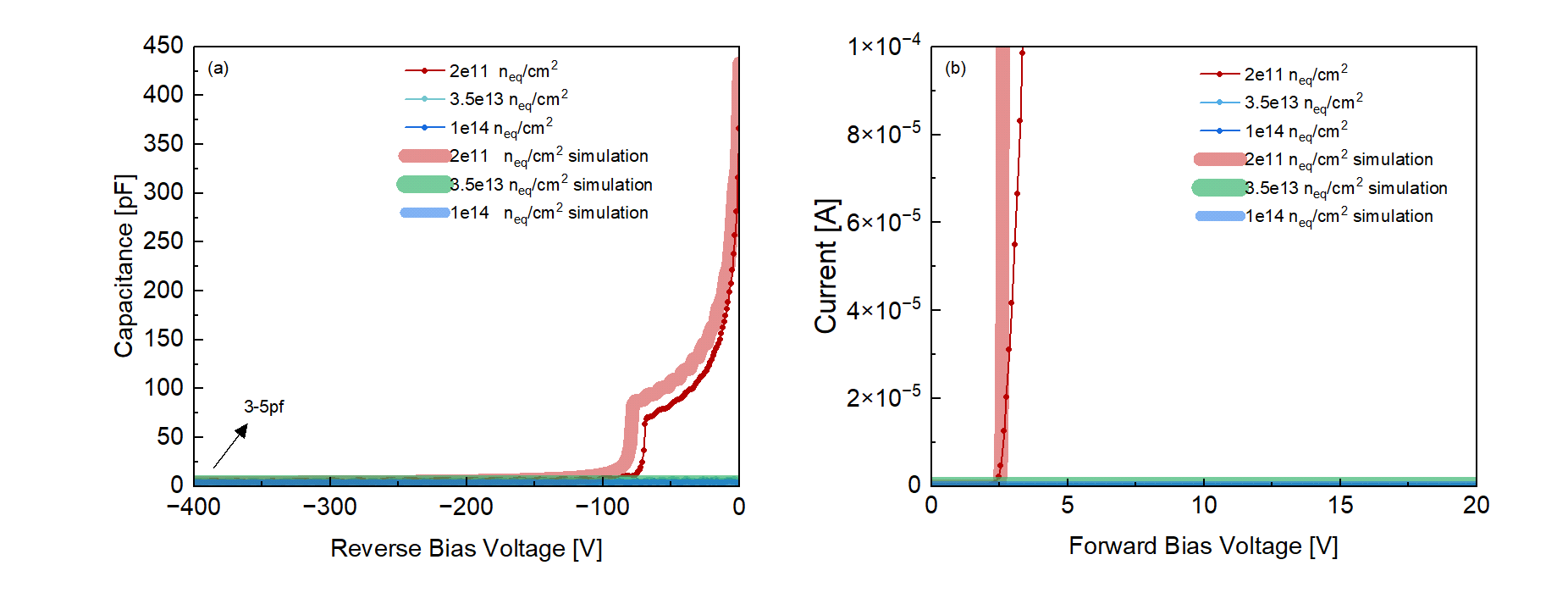}}

    \caption{(a) Capacitor characteristics test results and simulation results. (b) Forward current characteristics test results and simulation results.}
    \label{fig4}
\end{figure*}

 Based on this relation, the parameters extracted from experimental data were simulated and analyzed using TCAD software, with the results shown in \figurename~\ref{fig4}. The simulation results demonstrate well agreement with the experimental data, effectively validating the accuracy of the model. 
As the irradiation flux increases, the effective doping concentration tends to decay. This results in a depletion characteristic that is independent of the electric field and resembles the behavior of a parallel plate capacitor. In addition, under forward bias conditions, the injected minority carriers undergo enhanced recombination at deep-level defect centers, thereby resulting in an observable threshold voltage rise.\par

\begin{figure*}[!t]
    \centerline{\includegraphics[width=7in]{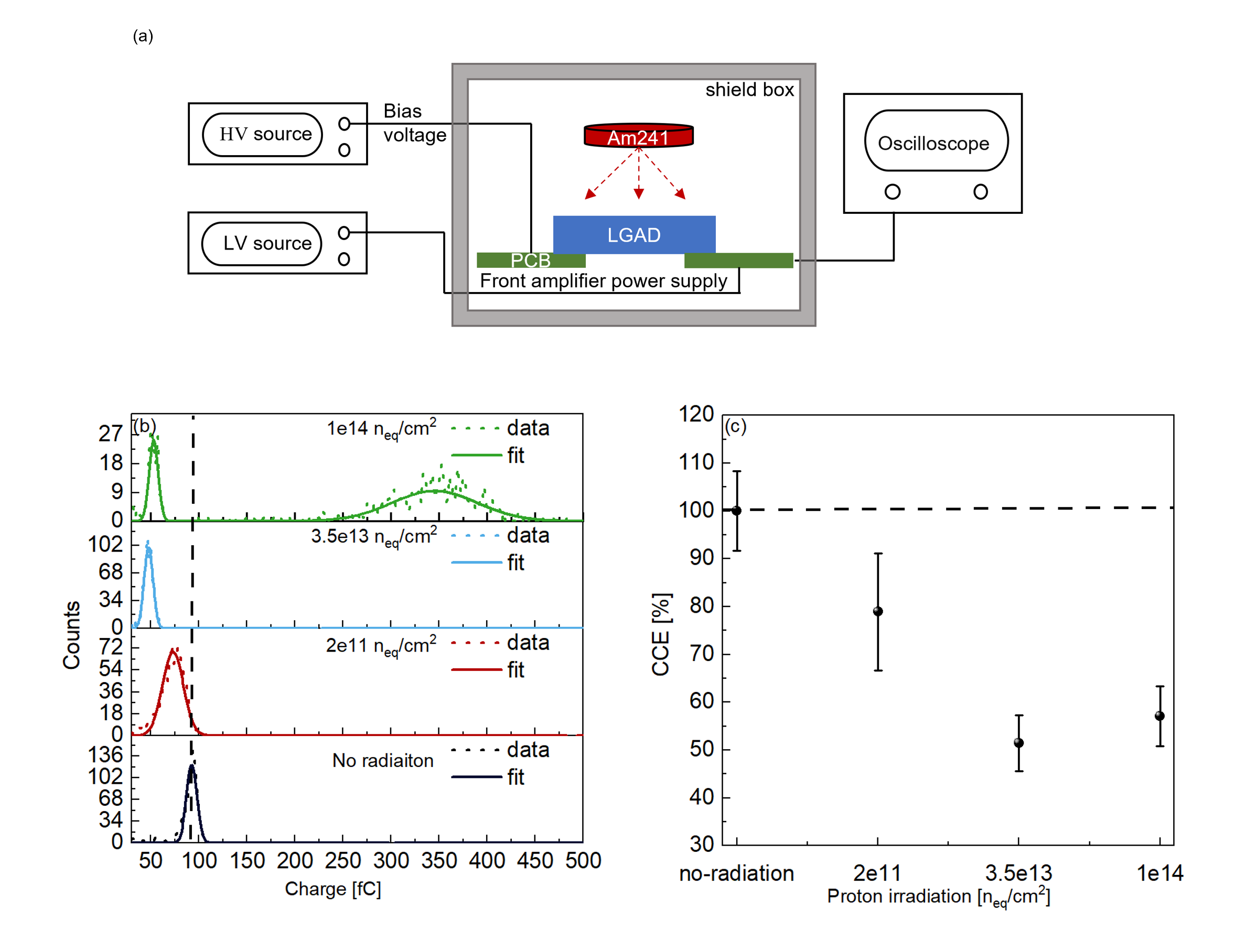}}

    \caption{(a) Experimental setup for $\alpha$ particle measurement. (b) The Gauss fit result of collected charge.(c) The~ CCE performance of different irradiated devices @300~V. The black dashed line indicates the 100\% charges collected in no-radiation LGAD}
    \label{fig5}
\end{figure*}

To further investigate the alpha-particle charge collection characteristics of 4H-SiC LGAD devices before and after proton irradiation, the experimental setup is illustrated as \figurename~\ref{fig5} (a). The setup includes $^{241}$Am radioactive source, 4H-SiC LGAD, single channel electronic readout board with SiGe amplifier,  high voltage source (keithley 2470), low voltage source (GPD-3303SGWINSTE) and oscilloscope (DPO-7354C, Tektronix10GHz). Given that the readout board employs a transimpedance amplifier for internal amplification, a signal generator was utilized to calibrate $R_f$.
Due to the high frequency of the 4H-SiC LGAD signal, a bandwidth of 2.5 GHz was utilised by the oscilloscope for the purpose of sampling. To analyse the charge collection, the signal area is divided by $R_f$ to convert voltage to current. Each set of data was fitted using a Gaussian distribution, with the collected charge being the centre value of the fit. The results of the fit and CCE are shown in \figurename~\ref{fig5} (b) and (c), respectively. The results demonstrate that at flux of 3.5$\times$10$^{13}~n_{eq}/cm^{2}$, there is a decrease in the amount of charge collected by the device. The amount of charge collected drops from 92.4~fC to about 47.5~fC, CCE drops to 51.5\%. When the flux reaches 1$\times$10$^{14}~n_{eq}/cm^{2}$, the amount of charge collected holds steady.\par 
Notably, at a flux of 1$\times$10$^{14}~n_{eq}/cm^{2}$, the detector exhibits Dual-mode charge collection behavior, characterized by two distinct charge collection of 52.7 fC (primary) and 300 fC (secondary). This phenomenon is hypothetically attributed to non-uniform electric field distribution within the device.  Combined with the DLTS results showing altered defect capture cross-sections, we hypothesize that the defects formed electric ligand-like structures analogous to an electric dipole.  The observed inhomogeneity in the electric field distribution within the device likely generated localized field enhancements exceeding the baseline magnitude, potentially triggering an increased rate of carrier multiplication within the specific channel. This effect resulted in a multiplication factor surpassing the initial theoretical predictions, a hypothesis currently under rigorous experimental validation to establish its validity and quantify the underlying mechanisms.\par

\section{Conclusions and outlook}

This study investigates the performance evolution of 4H-SiC LGAD devices under 80 MeV proton irradiation. XRD characterization confirms that the fully epitaxial-grown devices maintain perfect single-crystalline structure before and after irradiation. Key electrical characteristics include: (1) reverse leakage current decreases significantly by 2 to 4 orders of magnitude within the flux range of $2\times 10^{11} - 1\times 10^{14}~n_{eq}/cm^{2}$; (2) forward threshold voltage exhibits flux-dependent increase; (3) capacitance shows voltage-independent behavior at flux $ \geq 3.5\times 10^{13}~n_{eq}/cm^{2}$. DLTS analysis combined with TCAD simulation elucidates the influence of irradiation flux on the electrical characteristics of the devices. Remarkably, the maximum degradation of $\alpha$-particle charge collection efficiency is limited to only 48.5\%, maintaining performance comparable to unirradiated PIN detectors across the entire irradiation range. Furthermore, at an irradiation flux of $1\times 10^{14}~n_{eq}/cm^{2}$, we observed an anomalous dual-mode charge collection phenomenon. The underlying physical mechanism appears particularly complex and requires further systematic investigation.

\bibliographystyle{unsrt}
\bibliography{sample}

\end{document}